# Time-resolved Absorptance and Melt Pool Dynamics during Intense Laser Irradiation of Metal

Brian J. Simonds[1], Jeffrey Sowards[1*], Josh Hadler[1], Erik Pfeif[1†], Boris Wilthan[1], Jack Tanner[1], Chandler Harris[1], Paul Williams[1], John Lehman[1]

[1] National Institute of Standards and Technology, 325 Broadway, Boulder, CO 80305 USA
[*] Currently at NASA Marshall Space Flight Center, Huntsville, AL 35812 USA
[†] Currently at Johns Manville, 10100 W. Ute Ave, Littleton, CO 80127

## I. Abstract

High irradiance lasers incident on metal surfaces create a complex, dynamic process through which the metal can rapidly change from highly reflective to strongly absorbing. Absolute knowledge of this process underpins important industrial laser processes like laser welding, cutting, and metal additive manufacturing. Determining the time-dependent absorptance of the laser light by a material is important, not only for gaining a fundamental understanding of the light-matter interaction, but also for improving process design in manufacturing. Measurements of the dynamic optical absorptance are notoriously difficult due to the rapidly changing nature of the absorbing medium. This data is also of vital importance to process modelers whose complex simulations need reliable, accurate input data; yet, there is very little available. In this work, we measure the time-dependent, reflected light during a 10 ms laser spot weld using an integrating sphere apparatus. From this, we calculate the dynamic absorptance for 1070 nm wavelength light incident on 316L stainless steel. The time resolution of our experiment (< 1 µs) allows for the determination of the precise conditions under which several important physical phenomena occur, such as melt and keyhole formation. The average absorptances determined optically were compared to calorimetrically-determined values, and it was found that the calorimeter severely underestimated the absorbed energy due to mass lost during the spot weld. Weld nugget cross-sections are also presented in order to verify our interpretation of the optical results, as well as provide experimental data for weld model validation.

## II. Introduction

The recent accessibility of high-power laser systems has greatly increased their range of practical applications, in particular for manufacturing. What was once useful only in niche applications has now become an important industrial tool, in large part due to greatly reduced cost and operational complexity. These systems are highly sought by industry for processes like laser welding, cutting, and additive manufacturing for the benefits of increased productivity, faster weld speeds, design flexibility and reduced capital expenditures. However, the availability of industrial high-power laser systems has outpaced the knowledge of how the high irradiance beams they produce interact with a metal target. This interaction is complicated by the fact that the high-power laser beams are dynamically changing the target surface over a large temperature range and through multiple phase changes. Therefore, the quantification of a basic property like absorptance, which determines energy coupling and thereby underpins the predictions of any laser-matter interaction simulation, is incredibly difficult. In fact, due to a lack of precise knowledge of the dynamic energy coupling at high irradiance levels, energy absorptance is typically treated as a tuning parameter for mathematical models. The goal of this work is to provide high-accuracy input and output data allowing modelers to validate the physics of their simulations and to elucidate physical phenomena such as melting and keyhole formation with absolute, dynamic absorptance measurements.

In this paper, we study the interaction of high irradiance laser light on a solid metal target, which is most analogous to laser welding. It also provides a suitable stepping stone for understanding the related, but much more complicated, systems found in additive manufacturing where light is incident on a metal powder. Laser welding simulations have been under development for decades to improve weld procedures and to gain a better understanding of the joining process. These are complex computer models that solve differential equations typically for heat transport, material mechanics, fluid flow, and vapor dynamics in order to predict metallurgical effects. The long-range vision of these simulations is to eliminate the time and money associated with empirical process development such that a qualified weld can be produced the first time it is attempted – in other words, that they are predictive. At present, however, there are several roadblocks that limit the utility of these simulations in real-world process development.

Currently, a model is empirically "calibrated" to select the appropriate input parameters that yield agreement with measured results.[1] Common inputs for laser weld models are thermophysical, mechanical, and optical property data of the base material over a wide range of phases and temperatures. Ideally, a laser weld modeler would be able to find accurate property data in the literature for their material of interest. In practice, however, the published literature will, at best, provide a starting point for model calibration, which is even less true for new or exotic alloys that have not been extensively studied. Even in the case of well-established materials, acceptable composition tolerances within a particular alloy class can lead to drastically different properties. As just one example, thermophysical properties of austenitic stainless steels have been found to vary drastically in the literature: surface tension due to sulfur[2,3] and oxygen[3] content; thermal diffusivity[4] and linear expansion coefficient[5] due to Fe:Cr:Ni ratios. Values for optical properties like reflectance, emittance, and absorptance can vary even more widely as they are dependent on surface roughness and sample preparation[6–8]. In addition, the values used for model inputs are results culled from numerous different laboratories, with research spanning several decades, and using different experimental techniques on materials from different feedstocks which, in turn, result in increased ambiguity for the modelled system. As a result, the model calibration procedure requires its own time-intensive, iterative loop in parallel to that needed for empirical process development. And, as the available parameter space created by the range of process parameters is quite large, one could argue that less robust models could be tuned to give acceptable results while missing important physics of the internal weld process. This is an obvious hindrance to the goal of using such models to reliably predict results such as thermal history and weld microstructure.

The ideal approach to this problem is to provide a source of high-accuracy experimental input and output data that can be used to confirm a model's prediction. Towards this goal, we provide the measured dynamic coupling efficiency during a single spot weld under well-known, albeit simple boundary conditions. This is a critical input parameter for laser weld modeling for which there is scant data in the literature. Cross-section images of the weld nuggets are provided as a measured output against which a model can be validated.

The material used was a National Institute of Standards and Technology (NIST) standard reference material (SRM) for 316L stainless steel[9]. As a SRM, the composition is maintained and certified by NIST and is available for public purchase, which allows further material properties to be obtained as model demands require and as experimental techniques improve. While many aspects of the experiments performed here exist in the literature, the novelty of this work is that it ties well-characterized optical measurements and weld results to a standard material in a single resource. In addition, determining accurate uncertainties are a vital part of this work as these determine the range of acceptable outcomes for weld simulations. As a complete data set, this work can be used to validate the underpinning physics and calculations of a simulation. Once achieved, this validated simulation can be used with more confidence to predict results for a more complicated material system for which the modeler provides thermophysical property data.

Previous measurements of optical coupling efficiency can be generalized into two categories: those that gave an average value and those that measured the coupling dynamically, that is, as a function of time. The former has typically employed a calorimetry-based approach where the temperature rise of the sample was measured and related to the total heat input from the laser[10–13]. Measurements of dynamic coupling have been completed by a few researchers who utilized an integrating sphere to capture the light scattered during a laser weld[14–17]. We take a similar approach as the nature of the process is highly dynamic with rapidly changing coupling. Also, it does not require altering the sample by attaching probes, does not rely on knowledge of a heat capacity value (thus introducing additional uncertainties), and can be integrated to give an absorbed energy result as in calorimetry-based methods. To this latter point, we have measured the average coupling efficiencies using a dynamic approach and compared the results to those we obtained from calorimetry measurements and find that during keyhole welding, the calorimeter measurements require an energy correction term due to mass loss, which introduces additional assumptions and uncertainty. Furthermore, the time-resolved data in this work has revealed several physical phenomena during the laser welding process not encapsulated in a single, calorimetry-based average coupling value. These include the solid-liquid phase change, the onset of keyhole formation and its instability, as well as hydrodynamic behavior of the molten weld pool.

This paper is organized as follows. A thorough description of the material, experimental procedures, and analytical methods used is provided first. The weld cross-section dimensions and images are given for a range of energy densities resulting in conduction through keyhole welds. Next, the dynamic light scattering results are presented along with the determined dynamic absorptance. These are then compared with calorimetry-based measurements performed under nominally identical conditions. Last, a thorough discussion is presented on the phenomena revealed in the dynamic data.

### III. Materials and Methods

#### a. Sample Preparation

For these experiments, we used NIST standard reference material (SRM) 1155a[9], which is a 316L stainless steel with the composition listed in Table 1. Samples were precisely cut into discs 11 mm in diameter and 2 mm thick with a mass of 1.464 ± .020 g. They were then polished with 240 grit sandpaper to remove large-scale features, followed by wet-sanding with 2400 and 4000 grit sandpaper for 30 seconds each to give a mirror-like finish. Surface roughness measurements of 5 identically prepared samples showed an average root-mean square (RMS) roughness of 79 ± 20 nm. Figure 1, which shows optical profilometry data (a.) and an optical image (b.) of a representative sample, reveals that the polishing procedure leaves some micrometer scale scratches on the surface. The sample surfaces were cleaned with acetone and blown dry with compressed gas before a laser spot weld.

The use of an SRM is important to this work for several reasons. First, it is a feedstock of material whose composition has been verified and will remain the same for all samples within the stated measurement tolerances. Thermophysical properties, in particular thermal diffusivity[18], can vary as a function of composition even within the tolerances defined for a particular alloy class. Therefore, by using an SRM for these measurements we can assume that the thermophysical properties vary negligibly from sample to sample. Furthermore, any future thermophysical data collected for this particular SRM (1155a) can be referenced to the data in this manuscript for the purposes of modeling. Thus, as measurement techniques improve for difficult-to-measure properties, like surface tension and viscosity of the melt, for instance, if performed on this SRM, the data presented here will still be relevant.

*Table 1 - Chemical composition of NIST SRM 1155a.[9]*

| Element | Fe | Cr | Ni | Mo | Mn | Si |
|---|---|---|---|---|---|---|
| Mass % | (64.71 ± 0.12) % | (17.803 ± 0.099) % | (12.471 ± 0.056) % | (2.188 ± 0.015) % | (1.593 ± 0.060) % | (0.521 ± 0.017) % |
| Element | Cu | Co | W | V | P | C |
| Mass % | (0.2431 ± 0.0050) % | (0.225 ± 0.018) % | (0.0809 ± 0.0059) % | (0.0725 ± 0.0046) % | (0.0271 ± 0.0012) % | (0.0260 ± 0.0036) % |
| Element | Nb | Ti | N | As | Sn | S |
| Mass % | (0.0082 ± 0.0014) % | (0.0039 ± 0.0012) % | (0.0482 ± 0.0024) % | (0.007 ± 0.003) % | (0.0069 ± 0.0013) % | (0.0020 ± 0.0009) % |

b. Weld Cross-sections

Weld nugget dimensions were obtained by cross-sectioning samples mounted and polished in bakelite. The weld centers were found by progressively polishing towards the center with 240 to 800 grit SiC paper. In order to develop a contrast between the weld nugget and base metal, the specimens were etched with aqua regia (3HCl:1HNO$_3$). Images were obtained using an optical stereoscope. Cross-section widths were compared to the pre-mounted, top-down widths, and the polishing process was repeated until the two numbers agreed to within four percent. It was found that the image plane of maximum width sometimes varied from the image plane at which the maximum weld depth occurred. The spatial difference between these two image planes was on the order of 10 µm, and the difference in weld depth between these two planes was less than 6% in all instances. The cross-section images shown in this work are those at the plane of maximum weld width. However, all reported values for weld width or weld depth are the maximal values measured at any plane. Additional polishing and etching beyond these planes gave weld nugget dimensions that became smaller which ensured that the numbers reported here are at the maximums to within the uncertainty reported. Uncertainty due to repeatability was estimated from cross-sections of 6 nominally identically prepared samples. This resulted in a standard deviation in width and depth of 7% and 4%, respectively.

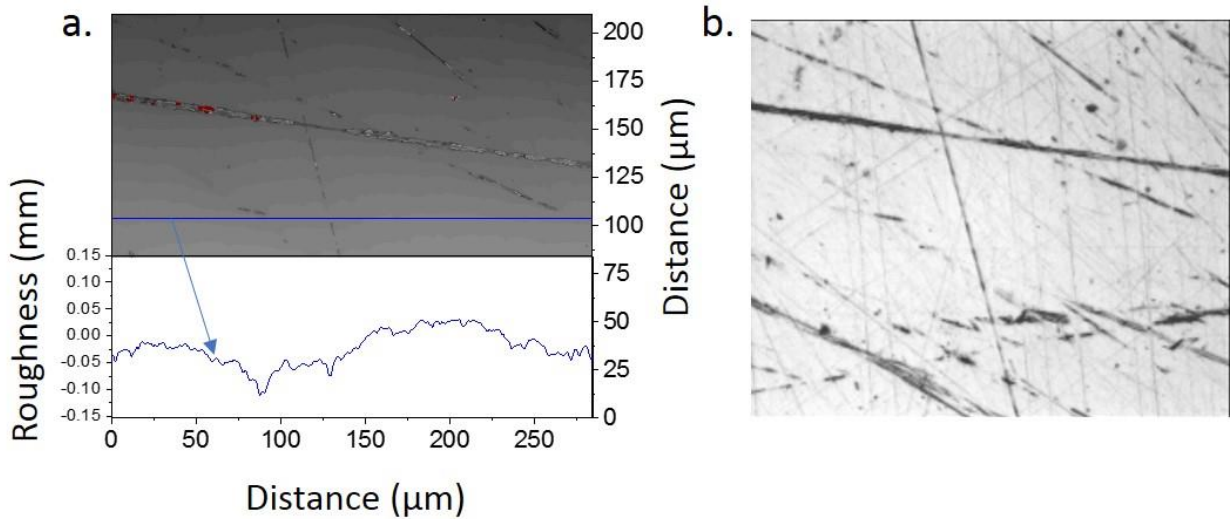

*Figure 1: a. Optical profilometry data and cross section of a sample representative of that used in this study. b. Optical microscope image of same location as used in a.*

c. Dynamic Absorptance Measurements

The general arrangement of the dynamic measurement is shown in Figure 2a. Laser spot welding was performed using a commercial fiber laser operating at 1070 nm with 10 ms illumination. The temporal profile of the laser output was roughly square with a rise-time of about 10 µs. Laser light was delivered to a commercial laser weld head by a 100 µm diameter, multimode fiber focused with a 150 mm focal length lens. This produced a top-hat spot at the focal plane that was 303 µm in diameter as measured by its full-width at the $1/e^2$ value (see Figure 2c.). The output energy was measured with a commercial power meter with a stated uncertainty of 3% before putting the sample in place.

An integrating sphere was used to capture the off-normal axis scattered light from the weld pool. The sphere was 3D-printed from a dark colored plastic in two halves as shown in Figure 2b. When combined, they had an outer diameter of 75 mm and inner diameter of 65 mm; the light entrance port was 7.5 mm in diameter. The target sample was placed on the opposite side of the sphere that had an aperture of 5 mm. The inner surface was coated with a $BaSO_4$ paint, which created a diffuse surface after the application of several (~20) thin layers. This coating was similar to that used in reference[19] and was particularly convenient for our purposes as imperfections caused by weld spatter could simply be painted over. A baffle was incorporated into the sphere design to ensure that neither the weld pool nor the plume was directly visible to the detection optics on the sphere wall (see Figure 2b).

Two photodiodes were used to detect the scattered light during the laser spot welds: one was attached to the laser weld head, and the other was fiber-coupled to the wall of the integrating sphere. Both were adjustable gain InGaAs photodiodes with 1070 nm bandpass filters preceding. The first was used to detect light scattered directly back into the weld head at normal incidence. This photodiode had a neutral density filter (optical density 2) and a rise time (based on the manufacturer's specification and its gain setting) of 21 ns. The second photodiode detected light from the integrating sphere and was passed through a 1070 nm bandpass filter. A second attenuation filter was used during the spot weld, which had a transmission at 1070 nm of 0.089 ± 1 %. Its gain setting corresponded to a response time of 450 ns.

For each measurement, normal incidence of the laser light was confirmed from the back reflection of a low-power guide beam coincident with the weld laser beam. The exact energy applied during a single spot weld was determined by averaging 10 pulses fired immediately before the spot weld was performed. The average standard deviation of these pulses was 0.4%. This process also allowed us to determine the amount of ambient light scattered within the weld held in the absence of a sample, which was subtracted from the weld head photodiode signal.

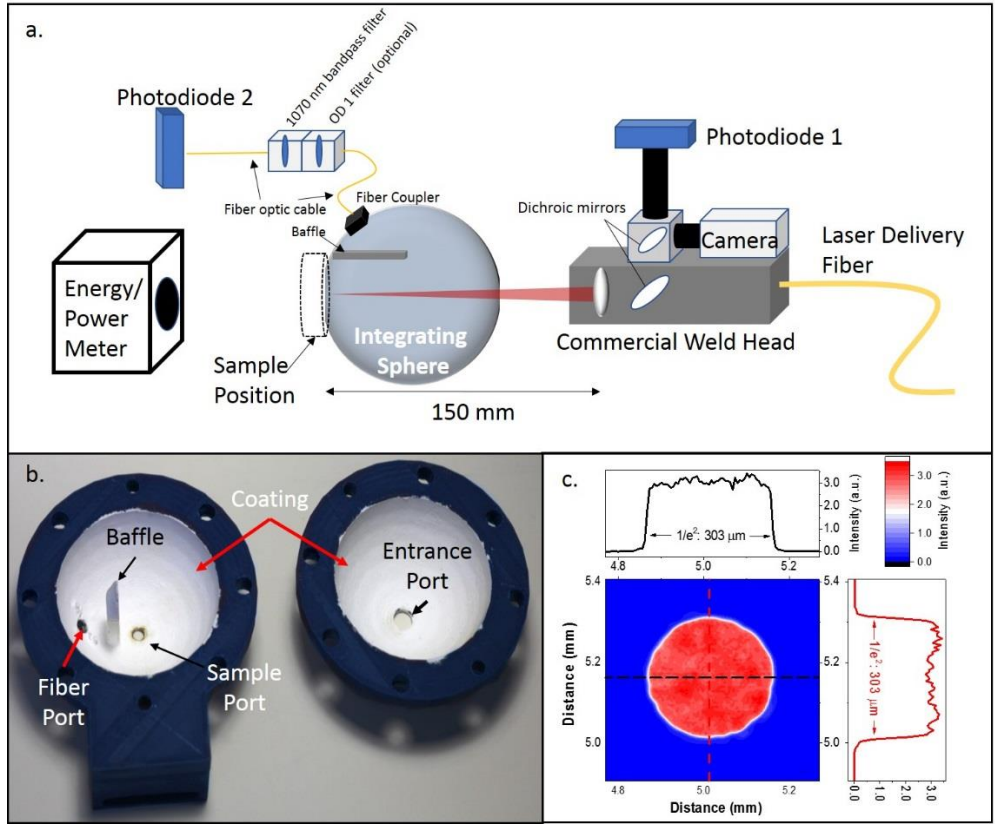

*Figure 2(color online) – The experimental setup of the optical measurements are shown in a. The inside halves of the 3D-printed integrating sphere with white BaSO$_4$ coating is shown in b. The measured beam profile at the focal point of the laser beam used is given in c.*

The measurement procedure for obtaining dynamic absorptance data is as follows. First, the weld head photodiode was calibrated by calculating the normal incidence reflected light from an optical-quality, fused quartz sample by subtracting the transmitted energy from the initial input energy. These values were determined by successive application of ten 10 ms pulses using a commercial energy meter, and assumes negligibly small absorption at 1070 nm in the quartz sample. This was performed over a range of input energies in order to determine a calibration constant, $C_{head}$ with units of W/mV, by a weighted linear fit. It was assumed that the calibration of the weld head photodiode was valid between successive spot weld measurements. Therefore, the absolute scattered power detected in the weld head, $P_{head}$, was determined from the raw weld head photodiode signal, $S_{head}$ (mV), according to

$$P_{head} = C_{head} * S_{head}. \tag{1}$$

A value for a single standard uncertainty in $P_{head}$ was determined through general propagation of errors.[20] The uncertainty in the calibration factor was determined by adding in quadrature the systematic errors resulting from the precisions of our oscilloscope (1%) and laser energy meter (3%) with the random errors determined from the deviation to the best linear fit. The random errors were minimized by averaging ten values for each data point in the fit. The result was a 3.7% standard uncertainty (confidence interval of 6%) in $P_{head}$.

The integrating sphere photodiode was calibrated separately before each spot weld as weld spatter could slightly alter the sphere's inner surface in successive runs. It was calibrated using a gold, diffuse scattering target in place of the sample in Figure 2a. A description of the target sample preparation and its measured bidirectional reflectance distribution function (BRDF) and reflectance losses can be found in reference[21]. A correction of 0.978 was applied to the measured signal to account for the losses due to sample absorption and from losses through the sphere entrance port. Due to the damage threshold limitation of this target, much lower energy densities were used for the calibration. To ensure that the signal range of our detection system covered the range that was typical during the laser spot weld, the attenuation filter was removed during the calibration. In a similar manner to the weld head photodiode, a calibration parameter $C_{sphere}$ was measured, and the absolute power $P_{sphere}$ determined from

$$P_{sphere} = C_{sphere} * \left( \frac{S_{sphere}}{T} \right). \quad (2)$$

The additional term $T$ is the transmission of the 1070 nm attenuation filter. Since the value of $C_{sphere}$ was determined before each spot weld, the standard uncertainty values of $P_{sphere}$ varied slightly, but remained between 3.5% and 4.0%.

The absorbed power, $P_{abs}(t)$, was found by subtracting the total light lost during the weld from the input power as

$$P_{abs}(t) = P_0(t) - \left( P_{head}(t) + P_{sphere}(t) \right). \quad (3)$$

The functional form of the time dependence of $P_0$ was determined from a measurement of the backscattered light from the quartz sample during the weld head photodiode calibration. It was normalized such that the integral of $P_0$ equaled the measured laser energy. Therefore, its uncertainty was the manufacturer's specification of the energy meter, or 3%. Normalizing $P_{abs}(t)$ by $P_0(t)$ gives the dynamic absorptance, $A(t)$, during the spot weld:

$$A(t) = \frac{P_{abs}(t)}{P_0(t)}. \quad (4)$$

The total energy absorbed, $E_{abs}$, during the spot weld was determined from the time integral of the absorbed power.

$$E_{abs} = \int P_{abs}(t) dt = \int \left( P_0(t) - P_{head}(t) - P_{sphere}(t) \right) dt. \quad (5)$$

If one assumes that the three components on the right-hand side of (5) are independent, then adding their uncertainties in quadrature gives the total uncertainty in $P_{abs}$, and therefore $E_{abs}$. An additional component of uncertainty was added in quadrature for the repeatability of the measurement of $E_{abs}$. This is determined from the standard deviation of five repeated measurements. The average coupling efficiency during the entire spot weld can be calculated according to

$$\eta_{coupling} = \frac{E_{abs}}{E_0} \quad (6)$$

with its uncertainty found by general propagation of errors.

   d. Calorimetry

Calorimetry was employed in an effort to validate the values of total absorbed energy from the optical measurements. The calorimeter was made of commercially available polystyrene, whose low thermal conductivity[22] makes it suitable for calorimetry. The sample rests in the base, which is approximately 5 x 10 x 10 cm³. A 1 cm thick polystyrene cover containing a 6 mm aperture is placed over the sample for a total calorimeter thickness of approximately 6 cm (see Figure 3a.). The sample surface was placed at the focus of the weld laser. The temperature rise of the sample was measured by two type-N thermocouples with leads 254 µm in diameter that were welded to the center and edge of the sample's rear surface. Representative thermocouple data is shown in Figure 3b. The energy collected by the calorimeter was calculated according to

$$E_0 = mC_p \Delta T_0 \qquad (7)$$

where $m$ is the puck mass and $C_p$ is the specific heat. The room temperature heat capacity of 316 stainless steel (0.50 J/g K)[23] was used as the temperature rise was only moderately above room temperature. Thermal transport simulations (not shown here) showed that the point where the sample came into contact with the polystyrene cover was never more than 14 °C above room temperature and thus did not damage the calorimeter.

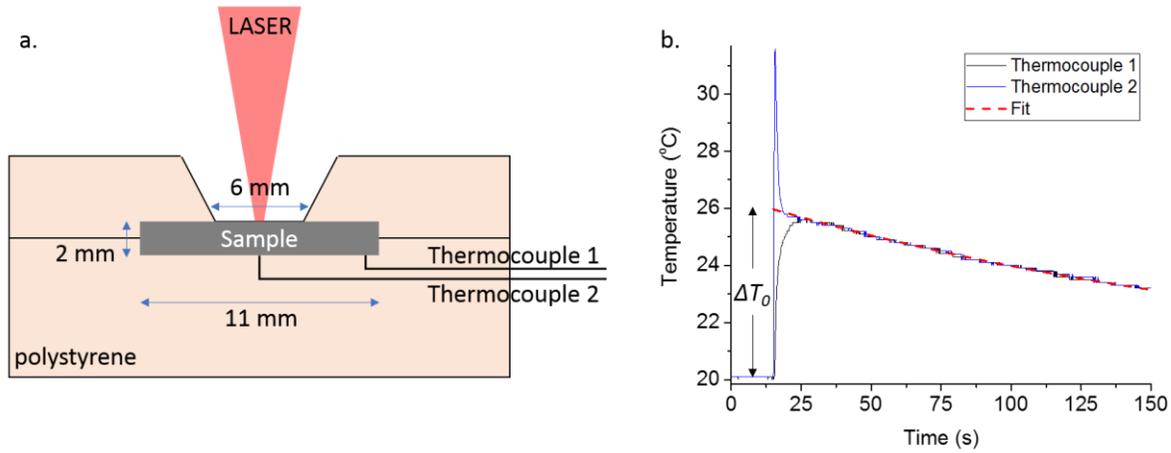

Figure 3 – Calorimeter design (a.) and representative data (b.).

For our calorimeter, as well as other similar devices[11,13,24], thermal losses are primarily from the front surface upon which the laser is incident. To account for these losses, the time regime where the two thermocouples values agreed was fit to an exponential decay curve from which the initial temperature rise ($\Delta T_0$) when the laser was fired was calculated (see Figure 3b.). This method, however, only corrects for those energy losses that are occurring when sample has reached equilibrium (about 5 s after the laser pulse). We consider two additional forms of energy loss that occur at much shorter times: 1) mass ejection during the spot weld, and 2) radiative losses from the keyhole during the laser pulse. It was found that these losses were quite significant, in particular during keyhole welding. Additionally, we made calculations of the radiative losses from the entire exposed surface between the time the laser is off to the time when the sample has equilibrated and found that they were not greater than 0.5% of $E_0$, and are therefore considered negligible.

Radiation and mass losses result only in a reduction to the total energy absorbed and therefore necessitated an energy correction term to be added to the calorimetry result. We estimated the energy lost from mass ejection according to

$$E_{vapor/liquid} = m\left(C_{p,s}\Delta T_1 + C_{p,l}\Delta T_2 + (H_m + H_v)T\right). \qquad (8)$$

The parameters are listed and defined in Table 2. The first term accounts for the energy needed to melt solid stainless steel and depends on the difference in temperature between room temperature and the melting point, $\Delta T_1$ = $T_{melt}$ - $T_{room}$. The next term describes the energy from heating the molten metal and the difference in temperature from the melting point to the vaporization temperature, $\Delta T_2$ = $T_{vapor}$ - $T_{melt}$. The final terms results from the phase changes to liquid and gas and the corresponding enthalpies of melting and vaporization, $H_m$ and $H_v$. The mass lost during a single weld was measured with a precision scale with a 100 ng readability by making 10 identical spot welds each on a fresh spot of a single sample and determining the average mass difference per shot. The use of equation (8) necessitates the presumption of the phase of the ejected mass: whether it is from ejected metal droplets or is in vapor form. As it is most likely that there is a combination of both, we consider the two extreme cases of all liquid droplets and all vapor as lower and upper bounds to the energy correction term. In the case of all liquid, the $H_v$ term is omitted from the calculation.

Table 2 – Parameters used for calorimetry energy correction terms.

| Parameter | Symbol | Value | Reference |
|---|---|---|---|
| Specific Heat, solid | $C_{p,s}$ | 0.50 (J/g K) | 23 |
| Specific Heat, liquid | $C_{p,l}$ | 0.79 (J/g K) | 23 |
| Enthalpy of melting | $H_m$ | 260 (J/g) | 23 |
| Enthalpy of vaporization | $H_v$ | 6360 (J/g) | 25,26 |
| Melting temperature | $T_{melt}$ | 1658 K | 23 |
| Vaporization temperature | $T_{vapor}$ | 3300 K | 25,26 |

Next, we consider the radiative energy lost from the weld pool region during the laser weld. This is determined by integrating the rate of energy radiated, $P_{radiative}$, from a surface during a 10 ms ($t_{pulse}$) weld:

$$P_{radiative} = \frac{E_{radiative}}{t_{pulse}} = \varepsilon \sigma T_{Weld}^4 A \qquad (9)$$

where σ is the Stefan-Boltzmann constant and $A$ is the surface area of the weld nugget defined by the measured weld diameter, $d_{weld}$. We approximate the infrared emissivity of the weld area, $\varepsilon$, during the weld as the average coupling efficiency, as this value represents an effective absorptivity over the duration of the spot weld. The temperature of the weld pool, $T_{Weld}$, must also be known, but is notoriously difficult to measure. Therefore, we were forced to make reasonable estimates based on literature values. We were aided in this by the work of Tan, *et al.* who simulated pulsed laser spot welds under very similar experimental conditions.[25] In this work, their "case 5" is roughly the same energy density as our $E_{in}$ = 3.46 J. Therefore, we use their predicted temperature of 3600 K for $E_{in}$ = 3.46 J and make a range of reasonable estimates for $T$ at higher and lower values of $E_{in}$.

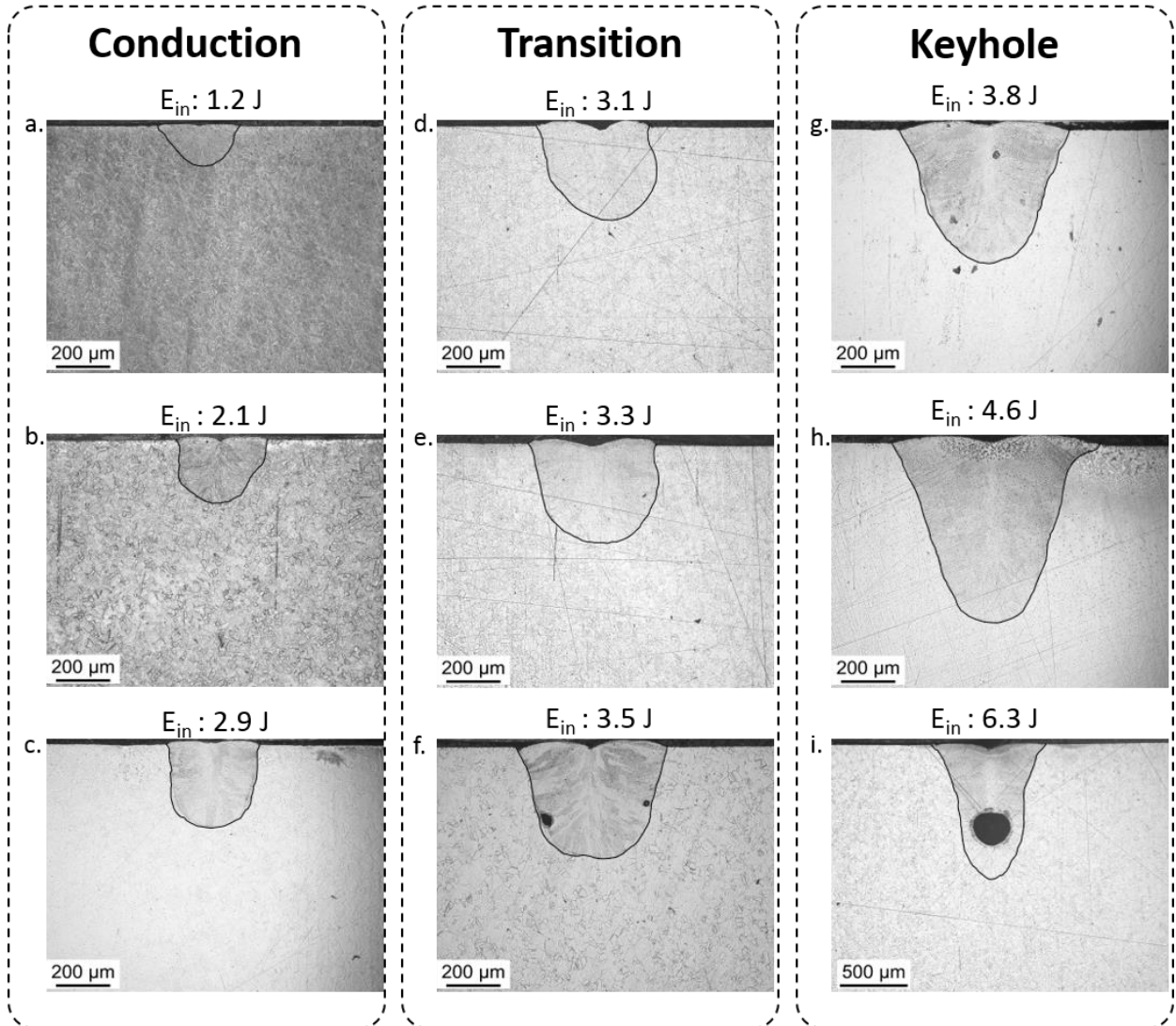

*Figure 4 – Laser weld cross-sections for 10 ms spot welds at the labeled input energy, $E_{in}$. They are grouped according to their appearance as either conduction, transition, or keyhole welds. Weld nugget outlines were traced with a dark line to improve visibility. Note that all have the same length scale, except for i. which is 2.5 times larger.*

IV. Results and Discussion

    a. Weld Cross-Section

Cross-section images of the spot welds are shown in Figure 4 with increasing input energy ($E_{in}$) from a. to i., and are grouped in columns according to their appearance as either conduction, transition, or keyhole welds. The lowest $E_{in}$ (a. - c.) clearly produces welds that are indicative of conduction welding (shallow, broad weld nuggets). Increased $E_{in}$ above about 3.5 J give the higher aspect ratio weld nuggets with fluted profiles that indicate keyhole welding (d. - f.). Intermediate weld nuggets share features of both and are labeled as "transition" welds. There is some subjectivity with respect to the exact classification of these welds, however, our optical measurements will make the distinction clearer. The measured weld nugget width ($W_{Weld}$) and length ($L_{Weld}$) are given in Figure 5a. The uncertainty in the width and length is the standard deviation of 6 separate spot welds performed at $E_{in}$ = 3.3 J. As the laser pulse energy is increased, the weld depth and diameter increase with a small inflection in the transition region from conduction to keyhole. Other studies of laser spot welding show similar increasing trends, but do not report an inflection[10,11,24]. The resolution of our data is undoubtedly improved as compared to these previous attempts due to the increased stability of our fiber laser source as opposed to the Nd:YAG sources used by those authors. This feature can more readily be seen in the aspect ratio ($L_{Weld}/W_{Weld}$) as shown in Figure 5b. The flattening of the aspect ratio is a sign of the transition from conduction and keyhole welds, which has also been observed previously[27].

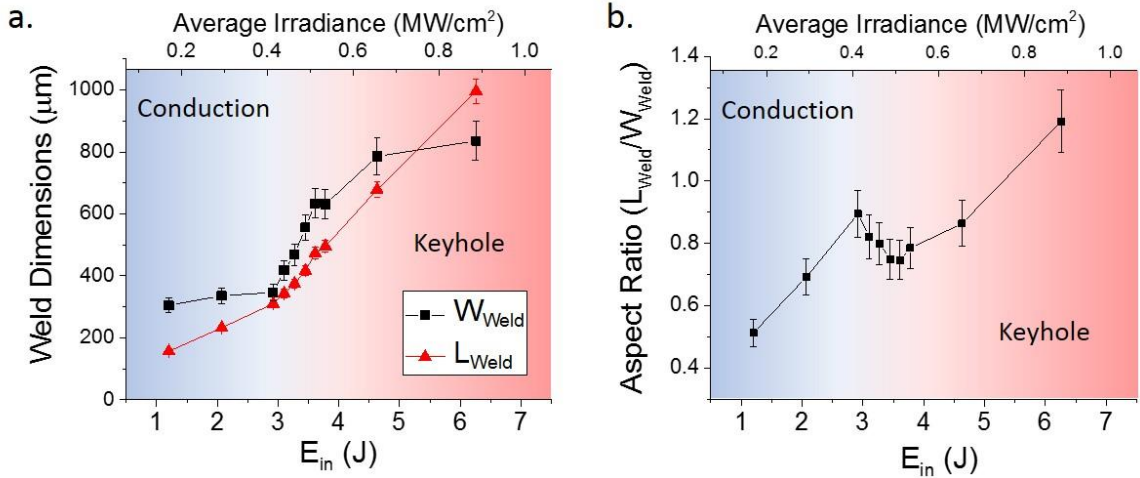

Figure 5 – a. Laser spot weld nugget depth and diameter as a function of input energy, $E_{in}$. (See also Table 3 for values.) b.: The aspect ratio ($L_{Weld}/W_{Weld}$) of the weld nuggets as a function of input energy.

b. Dynamic Absorptivity During Laser Spot Welding

Figure 6 shows the dynamic reflectance data from both photodiodes ($P_{sphere}(t)$ and $P_{head}(t)$; red and blue curves, respectively), as well as the power applied to the sample as a function of time ($P_0(t)$, black curves). These data have been calibrated to represent the absolute values for power. The figures have been organized according to their weld cross-section and coupling efficiency (discussed below) as either conduction (a. – c.), transition (d. - f.), or keyhole (g. - i.).

The data shown in Figure 6 show that, overall, very little light is returned to the weld head during spot welding and only contributes to the total scattered light at early times. This is consistent with a transition of the weld target from a mostly specular solid surface to a volatile molten pool.[16] It is also clear that as the input energy is increased this peak narrows, which is expected from more rapid melting. Therefore, at all but the lowest input energy, the absorptance calculation is almost entirely determined from light scattered within the integrating sphere.

The curves in Figure 6 are used to perform an energy balance calculation and compute the time-dependent absorptance using equations (3) and (4). This is plotted in Figure 7, which is organized from top-to-bottom in increasing $E_{in}$. There are several features in the absorptance data presented in Figure 7 that are worth discussing. Starting at short times below 1 ms, all data show a steep rise eventually followed by a decay. At low energy input, this happens at 600 μs and decreases rapidly as the energy input is increased. The sharpness of the feature suggests a dramatic change in the material properties like that which would accompany a phase change. Indeed, this feature can be readily explained as a result of melting with the following logic. The initial rise is due to solid state heating, which has been observed and credited to a variety of reasons. First, a rise in absorptivity with temperature is expected from a free-electron Drude model. However, the simplicity of the Drude theory only considers a perfectly flat surface. Observations of laser heating of steels under this condition show modest increases of absorptivity from about 0.09 to 0.12 over a temperature range of approximately 1000 °C from room temperature[28,29]. Oxide formation with heating has also been shown to increase absorptivity[30]. Kwon et al. attributed the doubling of absorptance (30 % to 60 %) during sub-melting laser heating at 1.06 μm of 304 stainless steel to oxidation[7]. Lastly, Wang et al. saw a similarly dramatic increase in absorptivity with surface temperature, which they ascribed to a semisolid sub-phase characterized by a spontaneous surface roughening[6].

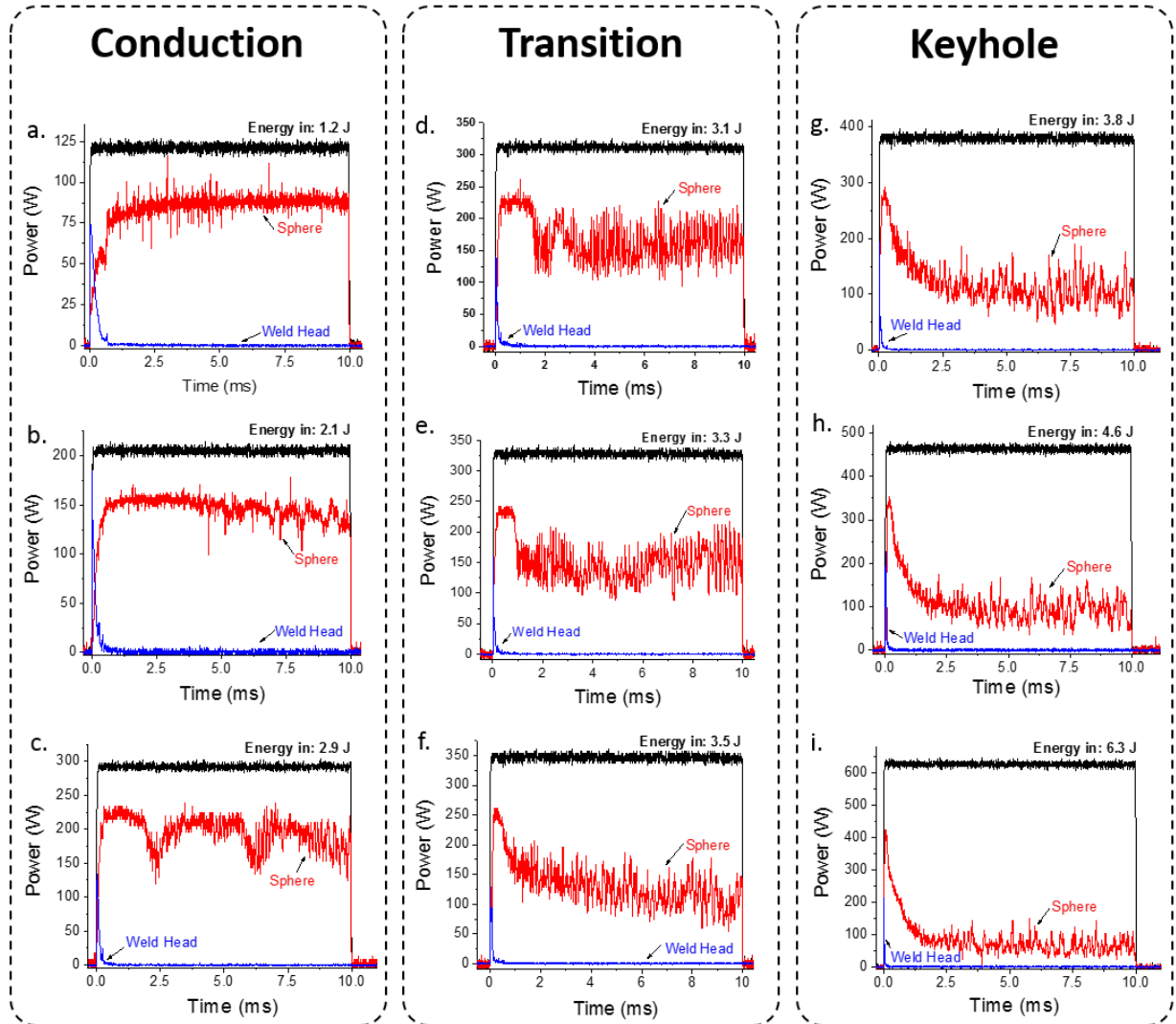

*Figure 6* – (color online) Scattered light results showing both the light scattered in the weld head ($P_{head}$, blue) and the integrating sphere ($P_{sphere}$, red) as compared to the power input ($P_0$, black).

The subsequent drop in absorptance can be explained by the surface roughness difference between the solid surface, which although polished is not optically flat, and that of a molten metal pool. Surface roughness has been shown experimentally to increase the absorptivity in steel alloys in the solid under a wide range of surface treatments[28,30,6,8]. Therefore, upon melting the surface tension of the liquid metal removes the small amount of roughness and presents a specular surface to the incoming laser beam.

This effect can easily be seen in Figure 6a. where an increase in the weld head signal around 700 µs is concurrent with a decrease in the sphere signal, resulting in an overall decrease in absorptance in Figure 7a. This explanation is counter to the expectation of a Drude model, which predicts a small step-function increase in absorptivity at the melting point due to an increase in free carriers[31]. However, based on our interpretation we believe that the effect is a surface roughness dominated and is therefore not considered in a Drude model. There is insufficient research in the literature on the effects of changes in surface oxidation at the melting point to laser absorption, but it seems reasonable to believe that this could also contribute along with surface roughness to the effects seen here.

The next feature visible in Figure 7c. - j. is a second rise in absorptance. As this feature only occurs in the materials where a keyhole geometry is seen in the cross-section images, it seems reasonable to believe that this rise in absorptance results from multiple-scattering due to keyhole formation. Especially telling is the fact that Figures 7c. and d., which are taken just at the onset of keyhole formation, show multiple absorptance "humps" that indicate that one is on the cusp of forming a stable keyhole. In fact, a threshold for stable keyhole formation as a function of absorbed laser power has been predicted theoretically with oscillations in and out of keyhole near the onset.[32] This assertion is also supported by the position of these data points on Figure 10 (discussed later) as they are at the initiation of the sharp transition region.

With both of these features, one can define characteristic times described as time-to-melt (Figure 8a.) and time-to-keyhole (Figure 8b.). (See also Table 3 for values.) The former is defined by the initial peak in absorptance. The latter we define as the time where the absorptance reaches 0.40, excepting the initial steep rise. This value was chosen as it lies on the rising edge of the feature that we have ascribed to keyhole formation. By integrating $P_{abs}(t)$ from time zero to the time-to-keyhole, one can calculate the amount of absorbed energy necessary for keyhole formation. This is also shown in Figure 8b. as the energy-to-keyhole. The same general trend exists for both in that the time to establish either is dramatically shortened with increasing pulse energy.

The time-to-keyhole presented in Figure 8b. shows an asymptotic behavior at higher energy. Fundamentally, it is expected that there be a lower limit to the time that it takes for keyhole formation as it is determined by the thermal transport properties, vaporization rates, and thermophysical properties of the material. A direct comparison to literature values for the time-to-keyhole is difficult given the dependence of results on specific experimental parameters (spot size, laser power distribution, materials, etc.). However, the order of magnitude we observe is confirmed by several modeling studies of keyhole formation.[25,26,33–35] Keyhole modeling of spot welding in 304 stainless steel with comparable parameters by Tan *et al.* calculated rapid keyhole growth on the order of milliseconds with tapering effects at 10 ms.[25] Modeling of Nd:YAG spot welding by Zhou *et al.* show keyhole formation by 6 ms, but have a 1 ms power ramp which would delay this compared to our experiments.[26] Lastly, a model for laser spot welding by Courtois *et al.* show keyhole formation times ranging from 2 ms to 8 ms over the range of parameters calculated.[33] Experimentally, an effort by Matsunawa *et al.* with high speed imaging of aluminum spot welding showed a time-to-melt of 200 µs and a time-to-keyhole of 1 ms.[36] Clearly these results support the general results we observe, however, future works will focus on gaining supporting evidence specific to our exact experimental arrangement.

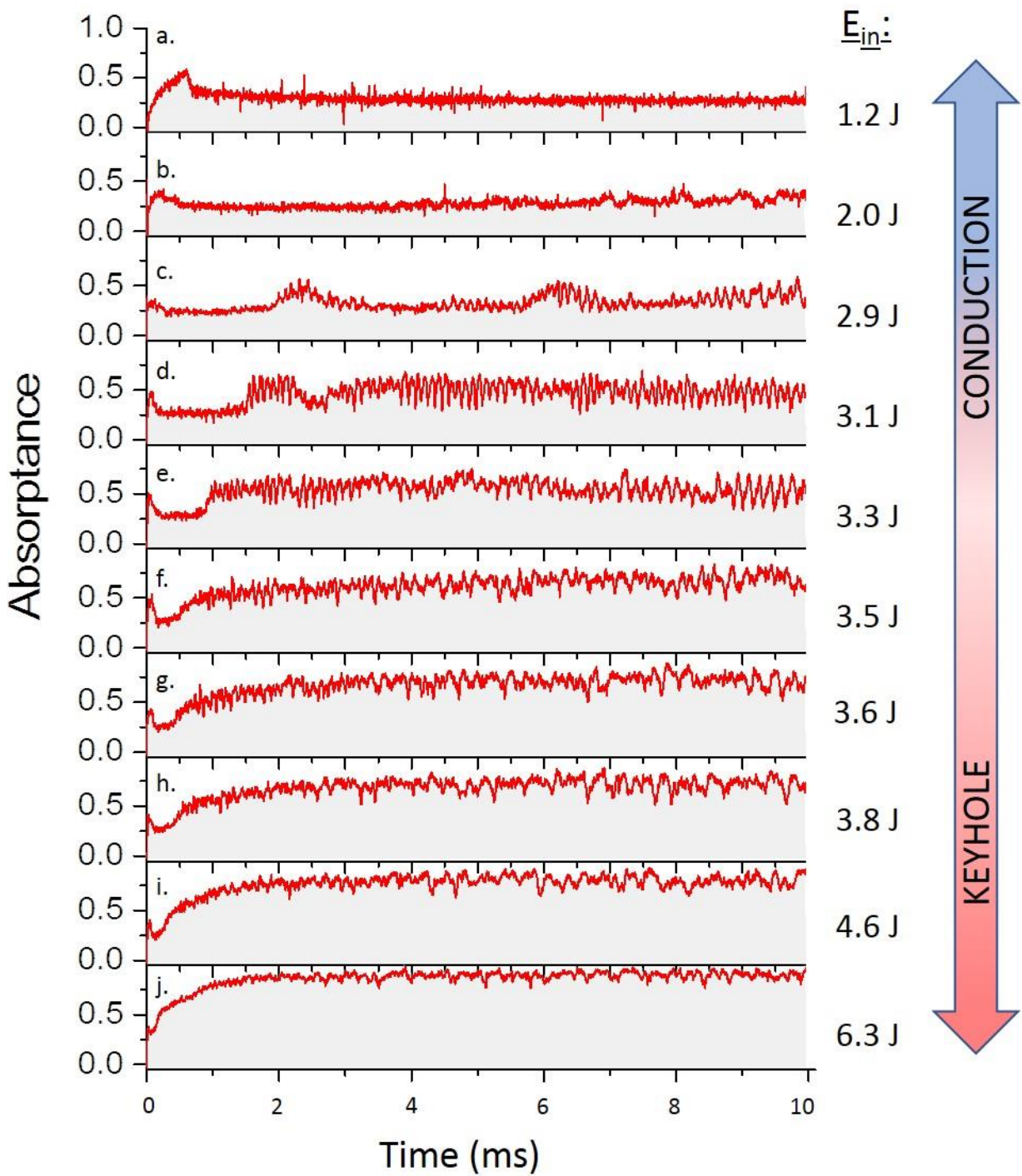

*Figure 7 (color online) – Dynamic absorptance data found from equation (4). The ratio of the shaded region to unshaded region in each graph is equal to the average coupling efficiency during the entire spot weld.*

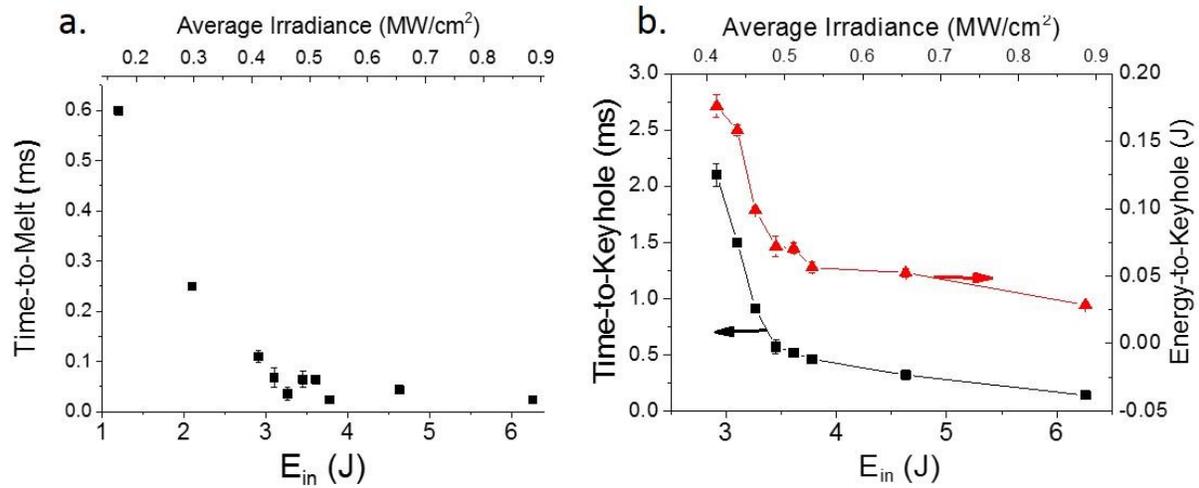

*Figure 8 – Values for the amount of time for the onset of melting (a.) and the formation of a keyhole (b.) as a function of laser pulse energy.*

The final feature of note in Figure 7 are the high-frequency, periodic oscillations that begin to appear in Figure 7c. (2.9 J) and are most intense in Figure 7d. (3.1 J) and Figure 7e. (3.3 J). These appear after the time-to-melt and only in those samples that eventually exhibit keyhole behavior. To further understand this behavior, the frequency components of these curves were calculated by a fast Fourier transform over the temporal region where they existed. These are plotted in Figure 9 on the same scale for easy comparison for curves in the conduction region (a.), transition region (b. and c.), and keyhole region (d.). In the frequency range of 0 to 15 kHz, the conduction result (a.) shows no discernable feature. In the transition region, strong frequency components appear between 7 and 10 kHz, shifting from higher to lower values as pulse energy is increased slightly. In keyhole mode (d.), the frequency components are much lower, in the few kHz regime.

Periodic oscillations associated with laser keyhole welding have been studied extensively over the past 30 years. An exact, quantitative comparison to literature results is difficult as they depend strongly on specific experimental conditions. A qualitative discussion is given here, with a more detailed analysis the subject of future work. Experimental evidence of oscillatory behavior has been seen from high-speed imaging[36–38], plume emission studies[39–42], acoustic emissions[40,43,44], and from scattered probe beams[37,45]. For laser welding near 1 μm wavelengths, where the plasma is not strongly absorbed by the plume, these oscillations are believed to be from mechanical oscillations of the molten metal. The frequency of oscillations typically observed are below 10 kHz but can vary widely depending of laser power[43], weld speed[40,42,43], spot size[39], shield gas flow[42], etc. These oscillations have been shown to correlate to weld penetration depth, and could therefore be used for *in situ* weld monitoring of weld quality[41,42,44]. Significant computational efforts have gone into calculating the oscillatory mode eigenfrequencies of the keyhole in radial, axial, and azimuthal directions[32,39,40,46], which result from instabilities around a balance of pressures that wish to push the keyhole open (ablation and gas flow) and those that wish to close it (surface tension, hydrostatic pressure, and fluid flow). These calculations generally agree with observations of oscillations below 10 kHz.

An interpretation of our data that is consistent with these previous results is as follows. The oscillations seen under keyhole formation conditions (Figure 9d.) appear to be most likely from keyhole dynamics of the type described and calculated by Klein et al.[32,39] to be in the 1-4 kHz regime. This has been verified experimentally by several groups[38–40,42–45]. These oscillations have a shorter wavelength and are less intense than those appearing at the onset of the transition region (Figure 7c. - e. and Figure 9b. - c.). An explanation for these oscillations, as well as the trend towards lower frequency over a very narrow input energy range, can be taken from Semak *et al.* who calculated the eigenfrequencies of a melt pool not yet in keyhole mode.[37] They did so for two different regimes: 1) those for surface capillary waves (drumhead oscillations) where the melt pool depth is smaller than the radius, and 2) volumetric oscillations where the depth is larger than the radius. The frequency of the latter was found to be 2-3 times less than the capillary waves as more molten mass is now oscillating. Therefore, the strong, high frequency oscillations at the onset of the transition region (Figure 9b.) are due to instability-induced surface waves resulting, most likely, from the initiation of surface vaporization.

A functional form for the natural frequency of capillary weld pool oscillations during gas tungsten arc (GTA) welding is given by Xiao and den Ouden[47] and depends only on surface tension, the density of the molten metal, and the weld diameter. Using values of 0.9 N/m for the surface tension in the presence of oxygen[23], 6900 kg/m$^3$ for the density of molten stainless steel[48], and the measured diameters from Figure 4, we find the values given by the vertical arrows in Figure 9 b. and c. The decrease in the intensity of these oscillations coincides with the deepening of the keyhole seen in Figure 4. This is a result as our absorption measurement being more sensitive to physical oscillations of the molten pool when it is in the plane of the original metal surface (as in conduction mode welding), which is why they are readily visible in Figure 7c. - e. Our optical signal would be less sensitive to the axial, radial and azimuthal oscillations that occur with a keyhole[32]. It is worth noting that although there are no fluctuations seen in Figure 9a. in the range presented, there are two relatively weak (around 1 kW) spectral components at 18 and 28 kHz, which is consistent of our interpretation of a small, relatively placid melt pool at low energies. Although these results are encouraging that our interpretation is plausible, we caution that future work is necessary to more definitely ascertain the exact nature and mode of these oscillations.

Table 3 – Optical results from integrating sphere measurements. Physical weld parameters of width ($W_{Weld}$) and length ($L_{Weld}$) are given, as well as the total absorbed energy ($E_{abs}$) and average coupling efficiency ($\eta_{coupling}$).

| $E_{in}$ (J) | Avg. Irradiance (MW/cm$^2$) | $W_{Weld}$ (µm) | $L_{Weld}$ (µm) | $E_{abs}$ (J) | $\eta_{coupling}$ | Time-to-Melt (ms) | Time-to-Keyhole (ms) |
|---|---|---|---|---|---|---|---|
| 1.2 | .170 | 304 | 156 | 0.378 | .31 | .60 | -- |
| 2.05 | .290 | 335 | 232 | 0.640 | .31 | 0.25 | -- |
| 2.92 | .412 | 346 | 310 | 1.020 | .35 | 0.11 | 2.1 |
| 3.10 | .439 | 418 | 343 | 1.550 | .50 | 0.068 | 1.5 |
| 3.27 | .462 | 468 | 374 | 1.794 | .55 | 0.036 | 0.91 |
| 3.45 | .489 | 556 | 417 | 2.209 | .64 | 0.064 | 0.57 |
| 3.61 | .511 | 633 | 472 | 2.535 | .70 | 0.064 | 0.52 |
| 3.78 | .535 | 631 | 495 | 2.620 | .69 | 0.024 | 0.46 |
| 4.63 | .655 | 786 | 679 | 3.580 | .77 | 0.044 | 0.32 |
| 6.26 | .889 | 836 | 996 | 5.370 | .86 | 0.024 | 0.14 |

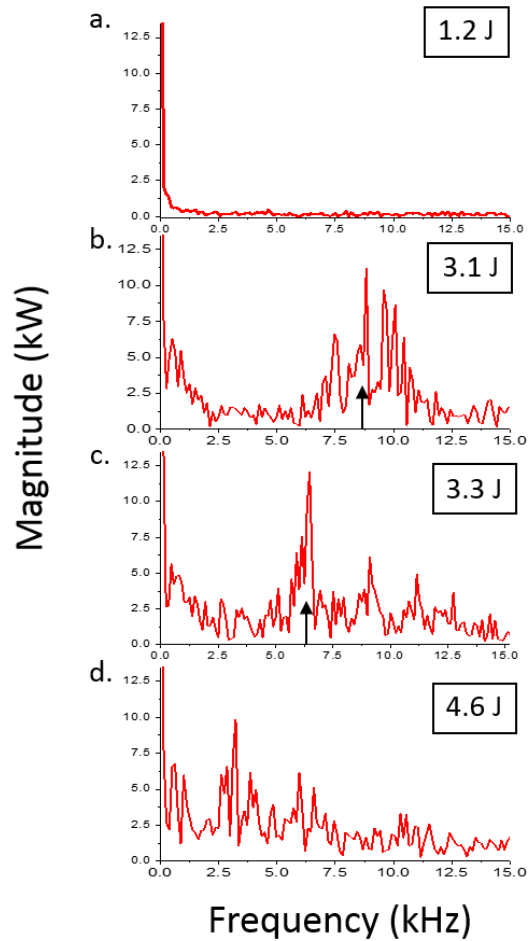

*Figure 9 - Frequency spectra of select dynamic absorptance data performed by fast Fourier transform. The vertical arrows in b. and c. give the calculated frequency of vibration of a melt pool as described in the text.*

c. Average coupling efficiency from optical and calorimeter data

The integral of the total absorbed power versus time is equal to the total amount of energy absorbed during the spot weld. From equations (4) - (6), this is proportional to the integrated area under the curves in Figure 7. The average weld coupling efficiency from optical data is plotted in Figure 10 as black squares. They are also given, along with other quantitative values from the integrating sphere measurements, in Table 3. At low values of $E_{in}$, the coupling efficiency remains roughly constant around 30%. A steep increase is seen after $E_{in}$ = 2.9 J when the efficiency is less than 35%, which by $E_{in}$ = 3.8 J has nearly doubled. At higher values of $E_{in}$, the efficiency continues to increase to nearly 90% for the highest value measured. This observation is consistent with many others who have seen a sharp increase in absorptance due to keyhole formation and the multiple reflections it allows.[10,12,13,24] Therefore, the sharp increase of efficiency can be ascribed as a transition from conduction to keyhole welding. This interpretation is consistent with the observations of the weld nugget cross-sections in Figure 4.

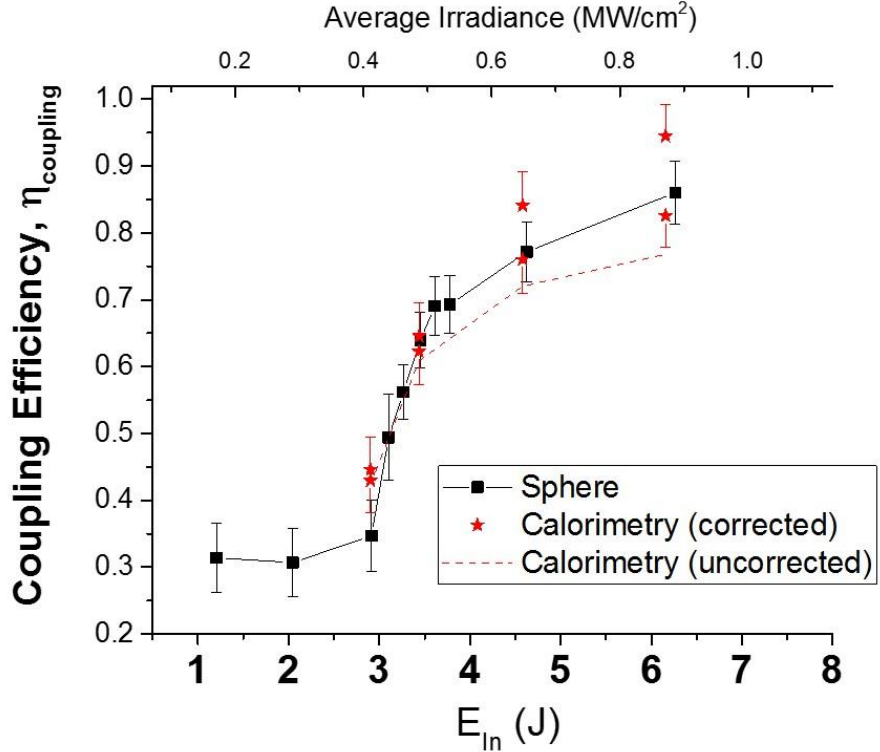

*Figure 10 – Average coupling efficiency, $\eta_{coupling}$, during a single, 10 ms laser spot weld as determined from optical data (black squares). The red stars represent upper and lower bounds of the coupling efficiency as measured by calorimetry and corrected for mass loss. The error bars in each are the standard uncertainty of the measurement. The dashed line shows the uncorrected calorimetry data.*

Calorimetry was used to validate the optical results as the final temperature rise resulting from the spot weld should correspond to the total absorbed energy through equation (7). This is also useful as it allows for comparison to similar experiments from the literature.[11–13] The coupling efficiency from calorimetry is simply the ratio of the absorbed energy ($E_0$) calculated from equation (7) divided by $E_{in}$. These results are plotted alongside the sphere results in Figure 10 with the uncorrected data given by the dashed line and corrected values as stars. Two data points for each $E_{in}$ are necessary to represent the upper and lower bounds for these losses as discussed previously. The uncertainties in these points reflect the standard uncertainties of the calorimeter measurement and not those of the assumptions for the corrections. Table 4 shows the measured temperature differences from calorimetry ($\Delta T_0$) and the absorbed energy, $E_0$, calculated from equation (7).

From Figure 10, at the $E_{in}$ = 2.91 J (that is, when in conduction mode), the calorimetry value is slightly larger than the sphere. However, the low value of $\Delta T_0$ (1.69 K) has a relatively high uncertainty such that the two values agree within a 95% confidence interval. As $E_{in}$ increases, the uncorrected average efficiency values from calorimetry begin to strongly deviate outside the range explained by experimental uncertainty (red dashed curve in Figure 10). This difference is accounted for by the energy losses not measured by the calorimeter, namely losses from vaporization, weld spatter, and radiation that we have attempted to quantify. The measured mass lost from each sample for a single spot weld is given Table 4, which is used in equation (8) to determine $E_{vapor/liquid}$ in Table 4. The radiative losses from the keyhole are determined from equation (9) using $W_{Weld}$ and $\varepsilon$, which are also listed in Table 4. Equation (9) requires that the temperature of the weld pool, $T_{weld}$, be known. As discussed previously, this was estimated from Tan et al.[25] with our estimates given in Table 4 along with the radiative energy correction term, $E_{radiative}$.

The right-most columns in Table 4 give lower and upper bounds for the total energy lost, $E_{lost}$. The lower bound assumes that all of the mass lost is in liquid form and the upper bound assumes it is in vapor form; both include $E_{radiative}$. These terms become the extremes of the correction values applied to our raw calorimetry data (red stars in Figure 10). The range of these bounds encompasses the integrating sphere result. At higher values of $E_{in}$, when keyhole formation is observed, $E_{lost}$ can be a significant fraction of the energy absorbed: a 23% correction to $E_0$ for the highest $E_{in}$ used in this work. Additionally, the magnitude of this correction relies on assumptions about the temperature and optical properties of the keyhole, as well as the phase of the mass ejected. The uncertainty in these parameters leads to a very large ambiguity in the actual absorbed energy as measured by calorimetry. Since the magnitude of the correction increases with the size of the keyhole, we consider the average coupling efficiency values from the integrating sphere to be much more robust as it does not suffer from these shortcomings. Other researchers have suspected as much[11,13], but until now a comparison to another technique has not been made. Nevertheless, we find good agreement between the optical results and those from calorimetry once the necessary corrections are made. However, we note that the corrections can be significant, especially during keyhole welding, and that the uncertainty involved with those corrections can render the results less reliable. Furthermore, our correction analysis does not consider unabsorbed light that is lost through scattering from ejected material. This scattered light would be collected and accounted for with the integrating sphere approach, but would further add to $E_{lost}$ in the calorimetry measurement. Quantifying this loss is beyond the scope of this work.

Table 4 – Results from calorimetry including energy correction terms for calorimeter results. See equations (7)-(9) for the definition of the terms.

| $E_{in}$ (J) | $\Delta T_0$ (K) | $E_0$ (J) | Mass loss (µg) | Emissivity, $\varepsilon$ | $E_{vapor}$ (J) | $E_{liquid}$ (J) | $T_{Weld}$ (K) | $E_{radiative}$ (J) | Total $E_{lost}$, bounds Lower (J) | Upper (J) |
|---|---|---|---|---|---|---|---|---|---|---|
| 2.91 | 1.69 | 1.23 | 5.8 | 0.35 | 0.0485 | 0.0116 | 3300 | 0.0026 | 0.014 | 0.062 |
| 3.44 | 2.87 | 2.10 | 12.6 | 0.64 | 0.1053 | 0.0252 | 3600 | 0.0122 | 0.047 | 0.127 |
| 4.58 | 4.52 | 3.30 | 58 | 0.77 | 0.4848 | 0.1160 | 4000 | 0.0513 | 0.183 | 0.552 |
| 6.16 | 6.48 | 4.73 | 115 | 0.86 | 0.9613 | 0.2299 | 4500 | 0.1052 | 0.355 | 1.087 |

## V. Conclusion

We have measured the dynamic absorptance during laser spot welding of 316L stainless steel NIST standard reference material with sub-1 µs time resolution. As a function of laser energy, we have unambiguously identified key features such as the time at which melting occurs and the keyhole forms. A sharp transition in the average coupling efficiency marks the transition from conduction mode welding to keyhole welding. This is confirmed with cross-section images of the laser weld nuggets. We associate with these measurements a rigorous uncertainty of energy absorption and laser irradiance. Furthermore, we have compared the value of average coupling efficiency obtained optically to those from calorimetry. We have found that the absorbed energy determined from calorimetry is severely underestimated due primarily to mass ejection during keyhole spot welding. We also observe temporal features during the weld evolution that may give us access to melt pool physics relying on viscosity and molten metal density.

VI. Acknowledgements

The authors are extremely grateful to Profs. Stephen Liu, Wenda Tan, and Wei Zhang for valuable discussions and insight. Matt Spidell is appreciated for his careful reading and analysis of the manuscript. We also wish to thank Cheryl Hawk for assistance with preparing the weld cross-sections.